\input harvmac

\lref\sem{G.W.Semenoff and R.J.Szabo,
  Int.\ J.\ Mod.\ Phys.\ A {\bf 12}, 2135 (1997)
  {\tt hep-th/9605140}}

\lref\rcollft{A. Jevicki and B. Sakita, ``Collective Field approach to the 
Large N Limit: Euclidean Field Theories," Nucl.Phys. {\bf B185} 89, 1981\semi
A. Jevicki and B. Sakita, ``The Quantum Collective Field Method and its Application 
to the Planar Limit," Nucl.Phys. {\bf B165} 511, 1980.} 

\lref\jevlev{A. Jevicki and H. Levine, 
``Large N Classical Equations and their Quantum Significance,"
Annals Phys. (N.Y.) {\bf 136}, 113 (1981).}

\lref\rode{R de Mello Koch and J.P. Rodrigues, 
``Systematic 1/N Corrections for Bosonic and Fermionic Vector Models without Auxiliary 
Fields," Phys.Rev. {\bf D54}, 7794 (1996). {\tt hep-th/9605079}}

\lref\cdp{M.Cavicchi, P. di Vecchia and I. Pesnado, 
``The Master Field of QCD in two dimensions and the 't Hooft Equation,"
Mod. Phys. Lett. {\bf A8}, 2427 (1993).{\tt hep-th/9306091}}

\lref\jevpap{A. Jevicki and N. Papanicoloau, 
``Classical Dynamics in the Large N Limit,"
Nucl.Phys. {\bf B171}, 362 (1980).}

\lref\bobt{Robert de Mello Koch, University of the Witwatersrand,
PhD Thesis, 1997.}
\lref\philt{Philippe A.F. Ferrer, University of the Witwatersrand,
PhD Thesis, 2005.}

\lref\dero{Robert de Mello Koch and Joao P. Rodrigues,
``Classical Integrability of Chiral QCD in two dimensions and Classical Curves,"
Mod.Phys.Lett. {\bf A12} 2445-2454, 1997, {\tt hep-th/9701138}.}

\lref\gn{D. Gross and A. Neveu, 
``Dynamical Symmetry Breaking in Asymptotically Free Field Theories,"
Phys. Rev. {\bf D10} 3235 (1974).}

\lref\papnev{A. Neveu and N. Papanicolaou,
``Integrability of the Classical Scalar and Symmetric Scalar - Pseudoscalar contact Fermi Interactions in two dimensions," 
Commun.Math.Phys. {\bf 58} 31, (1978)\semi
N. Papanicolaou, ``Pseudospin Classical Correspondence for Fermi Fields,"
Annals Phys. (N.Y.) {\bf 136}, 210 (1981).}

\lref\km{J.P. Keener and D.W. Mclaughlin, 
``A Green's Function for a Linear Equation Associated with Solitons,"
J. Math. Phys. {\bf 18}, 2008 (1977).}

\lref\ablowitz{M.J.Ablowitz, D.J. Kaup, A.C. Newell, H. Segur, 
``Method for Solving the Sine-Gordon Equation,"
Phys. Rev. Lett., 30, 25, 1262, (1973).}

\lref\gk{D. Gross and Y. Kitazawa, ``A Quenched Momentum Prescription for Large $N$ Theories,"
Nucl. Phys. {\bf B206} 440, 1982.}

\lref\pf{Phil Ferrer, work in progress.}

\lref\kt{M. Karowski and H.J. Thun, 
``Complete S Matrix of the O(2N) Gross-Neveu Model," Nucl.Phys. {\bf B190}, 61 (1981).}

\lref\kp{I.~R.~Klebanov and A.~M.~Polyakov,
  ``AdS dual of the critical O(N) vector model,''
  Phys.\ Lett.\ B {\bf 550}, 213 (2002)
  {\tt hep-th/0210114}.
}

\lref\jd{S.R.Das and A.Jevicki,``Large N collective fields and holography,''  Phys.Rev. D {\bf 68}, 044011 (2003)  {\tt hep-th/0304093}.}

 \Title{ \vbox{\baselineskip12pt\hbox{  Wits-CTP-023    }
}
}
 {\vbox{
\centerline{Fermionic Vector Model Solitons in the Large $N$ Limit}  }}

\centerline{ { Phil Ferrer}}

\vskip.1in 
\centerline{{\sl Center for Theoretical Physics,}}
\centerline{{\sl University of Witwatersrand,}}
\centerline{{\sl Wits, 2050  }}
\centerline{{\sl South Africa  }}
\vskip.1in 
\centerline{{\tt ferrerp@physics.wits.ac.za}}
\vskip .2in 
 
The large $N$ limit of fermionic vectors models is studied using bilocal variables,
in the framework of a collective field theory approach. The large $N$ configuration 
is determined completely using only classical solutions of the model. Further, the 
Bethe-Salpeter equations of the model are cast as a Green's function problem. One
of the main results of this work is to show that this Green's function is in fact
the large $N$ bilocal itself.

\Date{May 2005}

\newsec{Introduction}

Systematic large $N$ expansions remain among the most promising tools that can be used 
to probe the non-perturbative aspects of quantum field theories. For matrix theories,
the leading contribution is determined by summing all of the planar diagrams. Except
for the simplest models, this is still not possible. In contrast, the situation for 
vector models is much better (some work on vector/matrix models includes, for instance, \sem). 
Indeed, the large $N$ limit can often be constructed 
explicitely. The AdS/CFT correspondence relates the large $N$ limit of matrix theories
to a dual gravitational theory. One hopes, that with a better understanding of this
correspondence, large $N$ expansions of matrix theories will become possible. It would
be useful to have a simpler example with which the correspondence can be studied.
Recently, it has been suggested that the large $N$ expansions of vector models can
also be related to dual gravitational theories \kp, the so-called higher spin theories. 
In much the same way that vector models
were used to understand many of the features expected in QCD, it may be possible to
use the vector model/higher spin theories correspondence
as toy models for the AdS/CFT correspondence. Given this renewed interest in the
large $N$ limit of vector models, it is appropriate to revisit the outstanding
issues in the large $N$ expansions of these theories.

A concrete and very interesting proposal for a derivation of the 
vector model/higher spin theories correspondence was given in \jd. 
This article employed collective field theory to give a description of bosonic vector
models in terms of bilocal variables. The collective field theory is a promising tool
to attack this problem. Experience with the AdS/CFT correspondence suggests that 
supersymmetric examples may be the most amenable to analysis. For this reason,
the construction of the collective field theory for fermionic vector models is well
motivated. In particular, the descriptions of solitonic objects in the large $N$ limit
of fermionic vector models is not well understood. This was the primary
motivation for this work.

We discuss the description of the soliton of the Gross-Neveu model,
both at the leading and next to leading orders. Suprisingly, both are
determined largely by the classical solutions of the model. This is
likely to provide useful insights into how one would approach similar
questions for supersymmetric vector models which are relevant for the
vector model/higher spin theories correspondence.
Quite apart from the motivation provided by this correspondence,
the fluctuations at subleading order are interesting in their own right.
Indeed, they supply important information about the stability
of the solution (absence or prescence of tachyons) and about
the space of solutions (via the zero modes).

This paper is organized as follows: in section 2 the bilocal formalism 
is discussed and some central results stated.
Using the bilocal formalism, a leading order Ansatz is constructed in section 3,
first for the perturbative vacuum of the Gross-Neveu model, and then for the
soliton. Section 4 is concerned with the next to leading order fluctuations. 
In section 5, we discuss our results.

\newsec{The bilocal formalism}

We follow an approach based on collective field theory \rcollft\ 
to treat the large $N$ limit of vector models \jevlev, \rode. In this
approach a change in variables is implemented from the original fermionic 
field to time-ordered bilocals. The bilocal fields are singlets so that in
terms of these variables, all $N$ dependence is explicit. Consequently,
these are the correct variables in which to develop a systematic $1/N$ 
expansion. As usual, this change of variables gives rise to a Jacobian \rode. 
Incorporating the effects of this Jacobian produces an effective action 
that can be used to generate systematic corrections in $1/N$.

We start with an action invariant under $U(N)$ transformations. These
transformations act as ($\alpha$ is a spinor index)

$$\psi_{\alpha }\rightarrow \psi _{\alpha }^{\prime}=U\psi _{\alpha},
\qquad \psi _{\alpha }^{\dagger}\rightarrow \psi _{\alpha }^{\dagger \prime}
=\psi _{\alpha}^{\dagger}U^{\dagger},\qquad U\in U(N)$$

\noindent
on our fermionic variables.  We change variables from the original fermionc fields
to the invariant bilocals, given by

$$\sigma _{\alpha \beta}(x,y)=\bar{\psi}_{\alpha }^{a}(x)\psi _{\beta }^{a}(y).$$

\noindent
Requiring that the Schwinger-Dyson equations for arbitrary singlet operators
derived using the bilocal variables agree with the equations derived using 
the orginial variables yields a differential equation for the Jacobian $J$.
The equation can be solved exactly to obtain \rode

$$
\log J=-\left[ N+mL^{d}\delta ^{d}(0)\right]\Tr\log \sigma ,
$$

\noindent
where the trace runs both over the spinor indices and over functional space.
In the above expression, $m$ is the dimension of the Clifford algebra, $L^{d}$
is the volume of spacetime and the delta function is in momentum space.
The leading term in this expression was also obtained in \cdp. In terms of this
Jacobian the effective action is given by

$$\eqalign{
S_{eff} &=-i\log J + NS \cr
        &=iN \Tr\log \sigma +NS+imL^{d}\delta ^{d}(0)\Tr\log\sigma \cr
        &= NS_{0}+S_{1}.}$$

\noindent
It was shown in \rode\ that it is consistent to treat the last term as
subleading in a systematic $1/N$ expansion. In the above, the
fields have been rescaled such that an overall factor $N$ multiplies the
action. Consequently, the leading large $N$ configuration, $\sigma^0$, 
is now obtained by solving 

$$
\left.{\delta S_{0}\over \delta \sigma }\right| _{\sigma^0}=0.
$$

\noindent
The systematic expansion is now obtained by
expanding about this leading configuration

$$
\sigma_{\alpha\beta}(x,y)=\sigma_{\alpha\beta}^{0}(x,y)+{1\over\sqrt{N}}
\eta_{\alpha\beta }(x,y).$$

\noindent
Performing a systematic expansion of the effective action we obtain
 
\eqn\EffAct
{\eqalign{
S_{eff}&= N S_0(\sigma^0)
          +S_1(\sigma^0)-{i\over 2}C_2 +{1\over 2}D_2  \cr
&+\sum_{n=1}^\infty {1\over \sqrt{N^n}}\left(
i(-1)^{n+1}\left[{C_{n+2}\over (n+2)}
+ mL^{d}\delta (0){C_{n}\over n}\right]+{D_{n+2}\over (n+2)!}\right).  }}

\noindent
where

$$\eqalign{
D_n &=\int d^d x_1\cdots\left. \int d^d x_n \int d^d y_1 \cdots \int d^d y_n
{\delta ^{n}S\over \delta \sigma _{\alpha_1\beta_1}
(x_1,y_1)\cdots \delta \sigma_{\alpha_n\beta_n}(x_n,y_n)}
\right|_{\sigma^0} \cr
&\times \eta_{\alpha_1\beta_1}(x_1,y_1)\cdots\eta_{\alpha_n\beta_n}(x_n,y_n),}$$

\noindent
and

$$ C_{n}=\Tr( (\sigma^0)^{-1}\eta)^n . $$

\noindent
The first corrections to the leading configuration will come from the 
(quadratic in $\eta$) terms
$C_2$ and $D_2$. The use of these terms to generate 
corrections is one of the questions with which we will be concerned in this paper.
In particular, we solve the Bethe-Salpeter which is obtained as the wave 
equation for the bilocal following from the quadratic action. 

\newsec{Large N and the classical solution}

The results of the previous section (in particular, equation \EffAct) clearly
show that the effective action for the bilocal field is determined by the leading
large $N$ configuration $\sigma^0$. In this section, our aim is to argue that
this leading configuration can be constructed using the classical solutions of the
fermionic vector model. This extends earlier results obtained for bosonic
vector models \jevlev, \jevpap.
The result for fermionic models is based largely on the unpublished works 
\bobt, \philt. It extends and completes the discussion appearing in \dero.

As we mentioned in the introduction, our arguments are illsutrated using the
Gross-Neveu (GN) model. However, we would like to emphasize that the methods
developed in this article are general.
The Lagrangian of the GN model is given by \gn

$$
L=\bar{\psi}^a (x) \left(i\gamma ^\mu\partial_\mu\right)\psi^{a}
(x)+{1\over 2}g^2\left(\bar{\psi}^a (x)\psi^a(x)\right)^2 ,$$

\noindent
where $a=1,...,N$ is a color index. Note the presence of a quartic interaction 
and absence of a mass term. In the context of a path-integral quantization,
one would treat $\psi $ and $\bar\psi $ as Grassman valued fields. It is 
thus natural
to suspect that the relevant classical solutions are obtained by treating 
$\psi$ as a classical Grassman field. We will argue that this is not the case 
- one should treat $\psi$ as a normal commuting function.

Recall that for a vector model, the large $N$ limit is reached by holding 
$\lambda =g^{2}N$ fixed and taking $N\to\infty$. The leading bilocal 
configuration is determined by the Schwinger-Dyson equations. 
The simplification of the theory in the large $N$ limit is reflected
in the factorization of expectation values of singlets. The Schwinger-Dyson equations

$$
0=\int D\bar{\psi}D\psi {\delta\over\delta\bar{\psi}_{\alpha}^{a}(x)}
\left[ \bar\psi_\beta^a (y) e^{iS}\right],$$

\noindent
for the GN model are

$$ N\delta_{\alpha\beta}\delta^2 (x-y)\left\langle 1\right\rangle
 = i\left\langle\bar\psi_\beta^a (y ) i( \gamma ^\mu )^{\alpha\gamma}
\partial_\mu\psi _\gamma^a (x)
+ g^2\bar\psi_\beta^a ( y)
\psi_\alpha^a (x) \bar\psi_\rho^b (x) \psi_\rho^b (x)\right\rangle .$$

\noindent
In terms of the bilocal

$$\sigma_{\alpha\beta}(x,y)=\bar\psi_\alpha^a (x) \psi_\beta^a (y),$$

\noindent
the large $N$ Schwinger-Dyson equation\foot{That is, the Schwinger-Dyson equation
obtained after using factorization in the large N limit.} reads

$$
0=iN\delta_{\alpha\beta}\delta ^{2}(x-y)+i\left( \gamma ^{\mu }\right)
^{\alpha \nu }\partial _{x\mu }\sigma_{\beta \nu }(y,x)+g^{2}\sigma_{\beta
\alpha }(y,x)\sigma_{\rho \rho }(x,x),$$

\noindent
where the partial derivative is acting on $x$. To interpret this equation,
note that the {\it classical} equation of motion reads 

$$
0=i\left( \gamma ^{\mu }\right) ^{\alpha \nu }\partial _{x\mu }\sigma_{\beta
\nu }(y,x)+g^{2}\sigma_{\beta \alpha }(y,x)\sigma_{\rho \rho }(x,x).$$

\noindent
Thus, the large $N$ Schwinger-Dyson equation gives the classical dynamics
in the presence of a point source. We would like to emphasize 
that this result is
obtained {\it by treating $\psi $ as an ordinary commuting function} and
not as a Grassman valued field. Further, the point source does
{\it not} arise from the specific action used. It is a general result that the
large $N$ configuration of the bilocal singlet field is simply the classical
bilocal singlet field, obtained in the presence of a point source with strength
proportional to $N$ \bobt.

We now turn to the problem of explicitely determining the leading bilocal
configuration in terms of the classical solutions. It is simplest to illustrate
our method on the perturbative vacuum (no soliton) of the theory.
Towards this end, note that there are two linearly
independent classical solutions for any given energy

$$\eqalign{
\sigma_{\alpha\beta} 
(x,y) &=( \gamma ^{\mu }p_{\mu}+m )_{\alpha \beta }A_{i}^{2}
e^{-i(\omega_{i}(t_{x}-t_{y})-k_{i}(x-y))} \cr
\sigma_{\alpha \beta }( x,y) 
&=\left( \gamma ^{\mu }p_{\mu}-m\right)_{\alpha \beta }A_{i}^{2}
e^{i(\omega_{i}(t_{x}-t_{y})-k_{i}(x-y))},}$$

\noindent
where
 
$$
\omega _{i}^{2}=k_{i}^{2}+(g^{2}\sigma _{\alpha \alpha
}(x,x))^{2}=k_{i}^{2}+m^{2}.$$

\noindent
Using these two solutions, it is natural to make the Ansatz

$$\eqalign{
\sigma_{\alpha \beta}(x,y) &=\theta (t_{x}-t_{y})\left( \gamma ^{\mu
}p_{\mu }+m\right)_{\alpha \beta }A_{i}^{2}
e^{-i(\omega_{i}(t_{x}-t_{y})-k_{i}(x-y))}\cr
&-\theta (t_{y}-t_{x})\left( \gamma ^{\mu }p_{\mu }-m\right) _{\alpha \beta}
A_{i}^{2}e^{i(\omega _{i}(t_{x}-t_{y})-k_{i}(x-y))}.}$$
 
\noindent
It is clear that, for unequal times, this Ansatz solves the SD equation,
since then the SD equation reduces to the classical equation of motion.

Integrating the SD equation with respect to $t_{x}$ form 
$t_{y}-\varepsilon $ to $t_{y}+\varepsilon $ \bobt

$$\eqalign{
&-i\int_{t_{y}-\varepsilon}^{t_{y}+\varepsilon }dt_{x}N
\delta _{\alpha\beta }\delta (x-y)\delta (t_{x}-t_{y}) \cr
&=\int_{t_{y}-\varepsilon}^{t_{y}+\varepsilon}dt_x\left[ (i(\gamma
^{0})^{\alpha \nu }\partial _{t_{x}}-i(\gamma ^{i})^{\alpha \nu }\partial
_{ix})\sigma_{\beta \nu }(y,x)+g^{2}\sigma_{\beta \alpha }(y,x)
\sigma_{\rho \rho}(x,x)\right].}$$

\noindent
The expression on the left is an integral over a delta function and is
trivial to perform. Consider now the right side. It is only the first term
in the above expression, the time derivative, which contributes. We obtain

$$\eqalign{
-iN\delta _{\alpha \beta }\delta (x-y) &= i(\gamma ^{0})^{\alpha \nu }
\sigma_{\beta \nu }(y,x)\mid _{t_{y}-\varepsilon }^{t_{y}+\varepsilon } \cr
&=\sum_{i}\left[\theta (t_{x}-t_{y})i(\gamma ^{0})^{\alpha \nu }\left( \gamma
^{\mu }p_{\mu }+m\right) _{\beta \nu }A_{i}^{2}e^{-i(k_{i}(y-x))} \right.\cr
&\left. -\theta (t_{y}-t_{x})i(\gamma ^{0})^{\alpha \nu }\left( \gamma ^{\mu
}p_{\mu }-m\right) _{\beta \nu }A_{i}^{2}e^{i(k_{i}(y-x))}\right].}$$

\noindent
This last expression is rearranged to become

$$
-iN\delta _{\alpha \beta }\delta (x-y)=2\delta _{\alpha \beta
}\sum_{i}A_{i}^{2}\omega _{i}e^{-i(k_{i}(y-x))}.$$

\noindent
which can be solved to give 

$$
A_{i}=\sqrt{{N\over 2L\omega _{i}}}.
$$

\noindent
Substitution of 
$\omega _{i}^{2}=k_{i}^{2}+(g^{2}\sigma _{\alpha \alpha}(x,x))^{2}$ 
and the normalization condition $\sum_{i}A_{i}=1$ gives the constraint

$$
1={N\over L}g^{2}\sum_{i}{1\over \sqrt{k_{i}^{2}+(g^{2}\sigma _{\alpha
\alpha }(x,x))^{2}}}$$

\noindent
which is the gap equation for this theory \bobt. Although we have considered
the perturbative vacuum of the  field theory the argument is general. Indeed,
a completely parallel argument allows one to construct the large N bilocal
corresponding to the kink. Using the classical solution of \papnev\
we easily obtain

$$
\sigma _{\beta \theta }(y,x)=\sum_{k}[N\theta (t_{x}-t_{y})
\overline{\psi }_{\beta }(y,k)\psi _{\theta }(x,k)-N\theta (t_{y}-t_{x})
\overline{\psi }_{c\beta }(y,k)\psi _{c\theta }(x,k)],
$$

\noindent
where $\psi _{\theta }(x,k)$ is the classical solution of \papnev\ 

$$\eqalign{
\psi &=\left[ 
\matrix{ \psi _{1}(x) \cr \psi _{2}(x)} \right] =\left[ 
\matrix{
V^{1}(x)+iV^{2}(x) \cr 
U^{1}(x)+iU^{2}(x)}
\right] \cr
&=\left[ 
\matrix{
{\sqrt{\omega _{k}-k}\over g}
\left[-\tanh (mx)+{\omega _{k}-k+im\over 2\omega_{k}}\left( 1+\tanh (mx)\right) \right] \cr 
{\sqrt{\omega _{k}+k}\over g}\left[1-{\omega _{k}-k+im\over 2\omega _{k}}\left(
1+\tanh (mx)\right) \right]}
\right] e^{i(\omega _{k}t-kx)},}$$

\noindent
and $\psi_c$ is related to $\psi$ by charge conjugation. 
One can once again verify that this bilocal solves the large N SD equation
\bobt, \philt. Notice that this is in fact the solution to the classical Gross-Neveu
model equations with $N=1$. However, in the construction of the bilocal, it is clear
that the sum over the spectral parameter $k$ is playing the role of a sum over color.
This is reminiscent of the quenched momentum prescription for matrix models\gk. Exactly
the same behaviour was noticed even earlier in the context of bosonic vector 
models \jevlev, \jevpap.

\newsec{Fluctuations about the Leading Kink Solution}

In the previous two sections we have provided 
a construction for the bilocal representing the leading
large $N$ configuration. At finite $N$, corrections to the leading
bilocal will become important. In this section, we address the problem
of computing these corrections. The leading correction is determined
by the terms in the effective action that are quadratic in $\eta$.
These terms, when varied with respect to $\eta$ give the Bethe-Salpeter
equation of the theory, in the one soliton background. The result of 
this section is that we are able to solve this Bethe-Salpeter equation.
Once again, all that is needed for this construction is a knowledge of the 
classical solutions of the theory! It is primarily with the analysis
of this section in mind that we chose to study the Gross-Neveu model.
Using the fact that the model is integrable, we will be able to argue
that we have constrcuted a complete set of solutions to the Bethe-Salpeter
equation. Although our construction is completely general\foot{That is, as 
long as the leading configuration is known, our construction can be
carried out. Integrability of the model is not used; once the leading large
$N$ configuration is known the construction is (in principle) mechanical.}, 
the proof of the completeness of the solution set makes crucial use of the 
integrability of the model.

Our first observation is that the Bethe-Salpeter equation follows by
minimizing the quadratic action {\it or} by linearizing the Schwinger-Dyson
equations. Our use of the integrability of the Gross-Neveu model is to
use the scattering data as suitable parameters for this linearization.
Since the scattering data are a complete set, we are sure to obtain a complete
set of solutions to the Bethe-Salpeter equation. Our approach is motivated by
the stability analysis of \km.

The scalar field

$$ \sigma (x,t)=\bar{\psi}\psi ,$$

\noindent
can be used to define a new field $u(x,t)$ as

$$ \sigma (x,t)=e^{{i\over 2}u(x,t)}.$$

\noindent
Using the equations of motion for the Gross-Neveu model, 
one finds that $u(x,t)$ satisfies the Sine-Gordon equation. 
It is possible to explicitely obtain the dependence of $u(x,t)$
(and hence $\sigma (x,t)$) on the scattering data, using the methods
developed in \ablowitz. The field $\sigma (x)$ appears in the large $N$
Schwinger-Dyson equations as a ``potential"

$$ 0=iN\delta_{\alpha\beta}\delta ^{2}(x-y)
+i\left( \gamma ^{\mu }\right)^{\alpha \nu }\partial _{x\mu }
\sigma_{\beta \nu }(y,x)+g^{2}\sigma_{\beta
\alpha }(y,x)\sigma (x).$$

\noindent
Denote the generic scattering data parameter by $\Lambda$. Under 
$\Lambda\to\Lambda +\Delta\Lambda$, we obtain a second solution to the
Schwinger-Dyson equations. This follows as a consequence of the fact that
the Schwinger-Dyson equations have no expicit dependence on $\Lambda$.
It is a simple task to show, by expanding,

$$\sigma_{\alpha\beta}(x,y;\Lambda+\Delta\Lambda)=
\sigma_{\alpha\beta}(x,y;\Lambda)+
{\delta \sigma_{\alpha\beta}(x,y;\Lambda)\over\delta\Lambda}\Delta\Lambda,$$

\noindent
and using the fact that both $\sigma_{\alpha\beta}(x,y;\Lambda+\Delta\Lambda)$ and
$\sigma_{\alpha\beta}(x,y;\Lambda)$ satisfy the Schwinger-Dyson equations, that
${\delta \sigma_{\alpha\beta}(x,y;\Lambda)\over\delta\Lambda}$ satisfies the
Bethe-Salpeter equation. Now, for the problem considered here, since we know the
dependence of $\sigma (x)$ on $\Lambda$ (thanks to the connection to the Sine-Gordon 
model) we can compute the variation of $\psi^a_\alpha (x)$ using the (linearized)
equation of motion

\eqn\Lin
{\eqalign{
(\partial _{t}+\partial _{x})\delta \psi _{2} &=-g^{2}(\sigma \delta 
\psi_{1}+\delta \sigma \psi _{1})  \cr
(\partial _{t}-\partial _{x})\delta \psi _{1} &=g^{2}(\sigma \delta 
\psi_{2}+\delta \sigma \psi _{2}).}}

\noindent
The variation of the bilocal then follows

$$\delta \sigma_{\alpha\beta}(x,y)=
\delta\bar\psi_\alpha^a (x) \psi_\beta^a (y)+
\bar\psi_\alpha^a (x)\delta \psi_\beta^a (y).$$

\noindent
The solution to the system of equations \Lin\ can be written as

$$\left[\matrix{\delta \psi_1 (x)\cr \delta\psi_2 (x)}\right]
= \int_{-\infty}^\infty d^2 y \left[
\matrix{G_{11}(x,y) &G_{12}(x,y)\cr G_{21}(x,y) &G_{22}(x,y)}\right]
\left[\matrix{\delta \psi_1 (y)\cr \delta\psi_2 (y)}\right],$$

\noindent
where the Green's function $G_{\alpha\beta}(x,y)$ satisfies

$$i\gamma^\mu\partial_\mu G(x,y)+g^2\sigma (x)G(x,y)=\delta^{(2)}(x-y).$$

\noindent
The central observation of this section is that the above equation is nothing 
but the equation for the large $N$ bilocal and as a result, the above Green's
function is equal to the large $N$ bilocal configuration.

For the case of the Gross-Neveu model, it is possible to explicitely implement
this construction. Further, the construction of a complete set of solutions to 
the Bethe-Salpeter equation can be used to compute the propagator for the fluctuation
$\eta$ in the background of the soliton. This allows one to compute (for example)
the leading $1/N$ correction to the soliton mass which is known \kt. The specific
form of the solutions to the Bethe-Salpeter equation, the use of these equation to
determine the $\eta$ propagator and corrections computed using this propagator will be
given in \pf.

If the model was not integrable, the form of $\delta\sigma$ would have been unknown.
In this case the large $N$ bilocal configuration 
continues to furnish a suitable Green's function. Using this Green's function
one obtains an integral
equation that can be iterated in much the same way that one treats scattering problems
in non-relativistic quantum mechanics.
In this case, one obtains the analog of a Born approximation for the solutions of the
Bethe-Salpeter equation.

\newsec{Conclusion}
We have studied the large $N$ limit of fermionic vectors models using a collective
field theory approach. The large $N$ configuration was determined completely using
only classical solutions of the equations of motion, 
obtained by treating the fermionic fields
as ordinary commuting functions. For the case of the Gross-Neveu model, we have 
demonstrated that a complete set of solutions to the Bethe-Salpeter equation can be
constructed. 
The central insight is that the solutions
can be determined using a Green's function method, and further that this Green's
function is in fact the large $N$ bilocal configuration itself!
For more general theories, which will not be integrable, this insight will
provide an iterative approximation scheme for the solutions to the Bethe-Salpeter
equation. It is surprising that so much of the dynamics of a fermionic
quantum field theory can be captured by well chosen classical solutions.

Our analysis of the Gross-Neveu model
may now be extended to construct the propagator for the soliton, 
and compute, for example, corrections to quantities of interest which can
be read from the (known) exact $S$-matrix of the model \kt. This work is
in progress \pf. It would be interesting to extend this analysis to 
other completely integrable models.

For models which are not integrable, we have outlined a procedure to determine
an integral equation for the solutions to the Bethe-Salpeter quation. The kernel
of this integral equation is the large $N$ bilocal. This integral equation could be
iterated to obtain approximate solutions. The accuracy and convergence of this iterative
procedure is an interesting question which deserves to be explored.

\bigskip

I would like to express my gratitude to R. de Mello Koch and J.A.P.
Rodrigues for discussion aiding in the completion of this work.

\bigskip

\listrefs

\vfill\eject
\end